\documentclass[conference]{IEEEtran}
\IEEEoverridecommandlockouts
\usepackage{booktabs}
\usepackage{cite}
\usepackage{comment}
\usepackage{amsmath,amssymb,amsfonts}
\usepackage{algorithmic}
\usepackage{graphicx}
\usepackage{textcomp}
\usepackage{xcolor}
\usepackage{hyperref}
\usepackage{multirow}
\usepackage{graphicx}
\usepackage{subfig}
\usepackage[linesnumbered,ruled,vlined]{algorithm2e}
\usepackage{tikz}
\def\BibTeX{{\rm B\kern-.05em{\sc i\kern-.025em b}\kern-.08em
    T\kern-.1667em\lower.7ex\hbox{E}\kern-.125emX}}
\usepackage{balance}

\hypersetup{
  colorlinks=true,
  linkcolor=blue,
  citecolor=blue,
  filecolor=magenta,
  urlcolor=blue
}

\def\BibTeX{{\rm B\kern-.05em{\sc i\kern-.025em b}\kern-.08em
    T\kern-.1667em\lower.7ex\hbox{E}\kern-.125emX}}
\begin{document}

\title{ECG-Based Stress Prediction with Power Spectral Density Features and Classification Models\\

}



\author{\IEEEauthorblockN{ Md. Mohibbul Haque Chowdhury\IEEEauthorrefmark{1}, Nafisa Anjum \IEEEauthorrefmark{2}, and Md. Rokonuzzaman Mim\IEEEauthorrefmark{3}}
\IEEEauthorblockA{\IEEEauthorrefmark{1}\IEEEauthorrefmark{2}\IEEEauthorrefmark{3}Dept. of Electrical \& Electronic Engineering (EEE), \\Rajshahi University of Engineering \& Technology, Bangladesh}
\IEEEauthorblockA{Emails: efazc05@gmail.com, nafisaanjum9999@gmail.com, rokonuzzaman@vu.edu.bd}
}

\maketitle

\begin{abstract}
Stress has emerged as a critical global health issue, contributing to cardiovascular disorders, depression, and several other long-term illnesses. Consequently, accurate and reliable stress monitoring systems are of growing importance. In this work, we propose a stress prediction framework based on electrocardiogram (ECG) signals recorded during multiple daily activities such as sitting, walking, and jogging. Frequency-domain indicators of autonomic nervous system activity were obtained through Power Spectral Density (PSD) analysis and utilized as input for machine learning models including Decision Tree, Random Forest, XGBoost, LightGBM, and CatBoost. In addition, deep learning approaches, namely Convolutional Neural Networks (CNN) and Long Short-Term Memory (LSTM) networks, were directly applied to the raw ECG signals. Our experiments highlight the effectiveness of ensemble-based classifiers, with CatBoost achieving 90\% accuracy. Moreover, the LSTM model provided superior results, attaining 94\% accuracy with balanced precision, recall, and F1-score, reflecting its strength in modeling temporal dependencies in ECG data. Overall, the findings suggest that integrating frequency-domain feature extraction with advanced learning algorithms enhances stress prediction and paves the way for real-time healthcare monitoring solutions.
\end{abstract}


\section{Introduction}

Stress is a physiological and psychological response to demanding or threatening situations, and chronic high stress is associated with serious health issues including cardiovascular disease, depression, and impaired cognitive function \cite{mentis2024applications}. Early detection of stress is important for preventing these outcomes and improving overall well-being \cite{dalmeida2021hrv}. Wearable sensors and biosignals provide a way to continuously monitor stress during daily \cite{taskasaplidis2024review}. Among these, the electrocardiogram (ECG) and heart rate variability (HRV) derived from ECG have proven to be reliable indicators of stress, reflecting autonomic nervous system responses to stressors \cite{dalmeida2021hrv, szakonyi2021efficient}. Unlike self-reported assessments, physiological signals are not easily influenced by voluntary control, making them more objective markers of stress \cite{elzeiny2020stress, nkurikiyeyezu1910effect}. 


However, detecting stress in real-world settings is difficult because conditions like resting, sleeping, or walking alter physiological signals. Posture and activity changes can affect heart rate and HRV, leading to confusion in stress detection models. \cite{hongn2025wearable}. For instance, an increased heart rate may result from stress or physical movement, and distinguishing between the two requires contextual awareness. Few studies have incorporated lifestyle context into their stress prediction models.\cite{taskasaplidis2024review}, even though research shows that including context such as activity status (e.g., walking or resting) can improve prediction accuracy\cite{hongn2025wearable, taskasaplidis2024review}. This limitation motivates researchers to predict stress across varied lifestyle situations using ECG signals and to investigate which conditions most strongly affect stress levels.

To address these challenges, machine learning (ML) algorithms have been widely applied to stress classification using ECG and HRV features. Models such as Decision Trees and Random Forests perform well in physiological signal analysis, providing interpretability and robustness in noisy data. \cite{bobade2020stress, dham2021mental}. Ensemble-based boosting models, including CatBoost, XGBoost, and LightGBM, have also been applied to stress recognition. These methods exploit non-linear relationships in HRV features and achieve improved accuracy compared to individual classifiers \cite{dalmeida2021hrv, szakonyi2021efficient, mentis2024applications}. Such tree-based and boosting algorithms remain appealing in wearable applications due to their efficiency and explainability \cite{taskasaplidis2024review}.

With recent advances, deep learning (DL) methods have become increasingly prominent in stress detection research. Convolutional Neural Networks (CNNs) have been effective in extracting complex patterns directly from ECG signals. For instance, Kang \textit{et al.} applied a CNN--LSTM model to ECG recordings and achieved high accuracy in mental stress classification \cite{kang2021classification}. Similarly, Ramteke \textit{et al.} introduced an attentive CNN architecture that utilized ECG scalograms for detecting acute stress levels \cite{ramteke2025acute}. Other approaches have transformed one-dimensional ECG signals into two-dimensional image representations, enabling the use of pre-trained CNN models with transfer learning \cite{ishaque2022detecting, elzeiny2020stress}. LSTM-based models have also demonstrated strong capability in capturing temporal dependencies of HRV sequences, further enhancing predictive performance \cite{ramteke2025acute}. Recent studies have also applied foundation models and Transformers for stress recognition, showing the potential of advanced DL frameworks in this area.\cite{phukan2024sonic}. Deep learning enables end-to-end modeling with high accuracy, as shown by multi-class studies achieving strong performance across stress levels. \cite{mortensen2023multi}.

\begin{figure*}[htbp]
\centerline{\includegraphics[width=\linewidth]{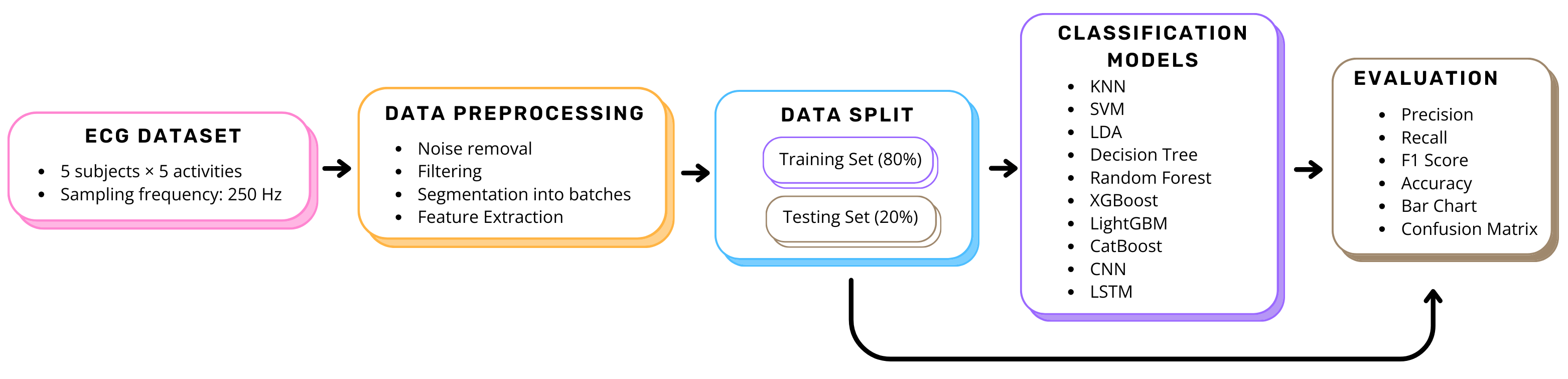}}
\caption{Flow chart illustrating the workflow of the proposed study.}
\label{fig:flowchart}
\end{figure*}

In this work, we extend prior research by predicting stress across lifestyle situations such as sleeping, walking, and sitting. We extract Power Spectral Density (PSD) features from ECG signals as frequency-domain markers of autonomic activity. These features are used to train ML models (Decision Tree, Random Forest, CatBoost, XGBoost, LightGBM) and DL models (CNN and LSTM). By comparing performance across contexts, we identify how activity affects stress prediction and highlight the strengths and limitations of each approach.


\section{Literature Review}

Early work on stress prediction used handcrafted features from ECG, especially heart rate variability (HRV), with traditional machine learning classifiers. HRV metrics from time and frequency domains, often estimated through Power Spectral Density (PSD), have proven reliable for stress detection.\cite{dalmeida2021hrv}. For example, Dalmeida and Masala \cite{dalmeida2021hrv} demonstrated that HRV features such as AVNN, SDNN, and RMSSD can achieve high recall in stress classification using models like Random Forest and Gradient Boosting.  

Szakonyi \textit{et al.} \cite{szakonyi2021efficient} studied acute stress detection with Ambient Assisted Living sensors and compared classifiers, finding ensemble methods like XGBoost and Random Forest performed best. They also showed that combining HRV feature domains improves accuracy over frequency features alone. Other studies confirmed that ensemble models and boosting algorithms improve predictive accuracy. Dham \textit{et al.} \cite{dham2021mental} reported that an ensemble stacking classifier combining ANN and SVM outperformed single models in stress detection tasks. Bobade and Vani \cite{bobade2020stress} compared classical ML models against a feed-forward neural network using multimodal physiological data, finding that Random Forest and SVM achieved strong performance on binary and multi-class stress detection.  

An important challenge addressed in the literature is the variability of stress responses across individuals. Nkurikiyeyezu \textit{et al.} \cite{nkurikiyeyezu1910effect} showed that generic models often fail to generalize, but incorporating person-specific calibration significantly improves predictive accuracy. Similarly, Elzeiny and Qaraqe \cite{elzeiny2020stress} highlighted the importance of personalization when classifying stress with frequency and spatial-domain images derived from PPG signals, showing that person-specific models greatly outperform generic ones.  

Recent reviews emphasize the need to consider lifestyle context when building wearable stress detection systems. Taskasaplidis \textit{et al.} \cite{taskasaplidis2024review} surveyed wearable-based stress detection methods and argued that incorporating contextual factors such as physical activity or posture is crucial for real-world use. Hongn \textit{et al.} \cite{hongn2025wearable} collected a dataset of wearable signals under both acute stress and exercise conditions, demonstrating that machine learning models such as XGBoost can successfully differentiate mental stress from physical exertion. These findings highlight the necessity of context-aware stress prediction in everyday life.

More recent work has turned to deep learning models to automatically extract features from ECG signals for stress prediction. CNNs and LSTMs have proven effective for capturing both spatial and temporal patterns in biosignals. Kang \textit{et al.} \cite{kang2021classification} proposed a CNN–LSTM architecture using ECG spectrograms, achieving over 98\% accuracy in mental stress classification. Ramteke \textit{et al.} \cite{ramteke2025acute} extended this line of work by introducing an attention-based CNN that processed ECG scalograms, reaching state-of-the-art accuracy for acute stress detection.  

Other researchers have focused on transforming ECG signals into image-based inputs. Ishaque \textit{et al.} \cite{ishaque2022detecting} converted ECG signals into two-dimensional images and applied transfer learning with pretrained CNNs, achieving strong accuracy while enabling model compression for wearable deployment. Elzeiny and Qaraqe \cite{elzeiny2020stress} similarly used CNNs on frequency-domain images derived from PPG, confirming the value of image representations for stress prediction.  

Deep learning has also been used for multi-class stress detection. Mortensen \textit{et al.} \cite{mortensen2023multi} developed a CNN model based on HRV features that successfully classified multiple stress levels with near-perfect accuracy. Phukan \textit{et al.} \cite{phukan2024sonic} proposed SONIC, an ensemble of vision foundation models applied to ECG signals, which achieved over 99\% accuracy on binary stress detection and high performance on multi-class tasks.  

Critical reviews point out that while deep learning shows high accuracy in controlled experiments, challenges remain in personalization and long-term monitoring. Mentis \textit{et al.} \cite{mentis2024applications} noted that AI systems for stress detection often reach about 90\% accuracy in research settings, but deployment requires addressing chronic stress and individual variability. These studies show the shift from ML to DL methods and highlight the need for personalized and context-aware approaches.

Overall, traditional ML methods such as Random Forest and XGBoost are effective for HRV and PSD-based stress prediction, but their performance is limited by individual and contextual variability. DL models including CNN, LSTM, and hybrids push accuracy further through automatic feature learning. Yet lifestyle context remains underexplored. Building on this, the present work evaluates both ML and DL models for stress prediction across daily situations such as sleeping, sitting, and walking using ECG-derived PSD features.

\section{Methodology}

This study follows a structured workflow to build stress prediction models from ECG signals. The process begins with dataset preparation and signal preprocessing, followed by feature extraction through Power Spectral Density (PSD) analysis. The extracted features are used in machine learning models, while deep learning architectures are applied to capture temporal and spatial patterns in the signals. Finally, an evaluation protocol is employed to assess model performance across lifestyle situations.

\subsection{Dataset}

This study uses ECG recordings obtained from three publicly available databases on PhysioNet. Five subjects were selected, and for each subject, two-minute recordings were collected for five activities: sitting, jogging, handbike driving, walking, and performing a mental arithmetic test. These activities were chosen to represent both resting states and stress-inducing conditions, allowing for the evaluation of stress prediction under varied lifestyle situations. All ECG signals were sampled at 250~Hz, providing sufficient resolution for accurate heart rate variability and frequency-domain analysis. Together, these datasets offer a reliable foundation for assessing stress detection models across physical and cognitive tasks.

\subsection{Preprocessing}

ECG recordings often contain noise that reduces signal quality and complicates feature extraction. Common artifacts include power line interference, motion artifacts, and baseline drift. Baseline drift, caused primarily by body movement and respiration, appears as a low-frequency fluctuation in the signal. To reduce these effects, filtering techniques were applied, including a high-pass filter and additional combined methods, to suppress baseline wander and improve clarity. Each ECG recording was then divided into smaller segments to facilitate analysis. The signals were partitioned into 100 equal batches, with each batch containing a fixed number of data points. This segmentation reduced computational load and ensured that the signals could be analyzed in consistent windows, preparing them for feature extraction and classification.

\begin{table}[!htbp]
\centering
\normalsize 
\caption{Description of Extracted Features}
\label{tab:feature_description}
\renewcommand{\arraystretch}{1.25}
\resizebox{\linewidth}{!}{%
\begin{tabular}{|l|p{9cm}|}
\hline
\textbf{Feature} & \textbf{Description} \\ \hline
No. Beats/min & Number of heartbeats per minute (derived from BVP/ECG). \\ \hline
HR (bpm) & Average heart rate in beats per minute. \\ \hline
SDNN (sec) & Standard deviation of NN intervals (overall HRV measure). \\ \hline
RMSSD (sec) & Root mean square of successive differences between NN intervals (short-term HRV). \\ \hline
NN50 (sec) & Count of successive NN interval differences greater than 50 ms (parasympathetic activity). \\ \hline
Stress Index (mv/sec$^2$) & Indicator of sympathetic activity; higher values indicate stress. \\ \hline
VLF & Very Low Frequency power component of HRV spectrum (0.003–0.04 Hz). \\ \hline
LF & Low Frequency power component (0.04–0.15 Hz), reflects both sympathetic and parasympathetic activity. \\ \hline
HF & High Frequency power component (0.15–0.40 Hz), reflects parasympathetic (vagal) activity. \\ \hline
LF to HF ratio & Ratio of LF to HF, indicator of sympathovagal balance. \\ \hline
PNS Index & Index representing parasympathetic nervous system activity. \\ \hline
SNS Index & Index representing sympathetic nervous system activity. \\ \hline
SNS to PNS ratio & Ratio of SNS to PNS indices, stress indicator. \\ \hline
CAB & Cardiac Autonomic Balance (relationship between SNS and PNS activity). \\ \hline
VSE & Vagal-Sympathetic Evaluation (quantifies autonomic function). \\ \hline
Kurtosis & Statistical measure describing the peakedness of the signal distribution. \\ \hline
Skewness & Statistical measure describing the asymmetry of the signal distribution. \\ \hline
\end{tabular}
} 
\end{table}

\subsection{Feature Extraction}

After preprocessing, a set of features was extracted to capture stress-related patterns in the ECG signals. 
R-peak detection was first performed to identify heartbeat locations and derive RR intervals, which serve as the basis for heart rate variability (HRV) analysis.  From these intervals, both time-domain and frequency-domain features were computed. Time-domain measures such as SDNN, RMSSD, and NN50 quantify short- and long-term fluctuations in heart rate, while additional indices like the Stress Index, PNS Index, and SNS Index provide markers of autonomic nervous system activity. Frequency-domain analysis was conducted using Power Spectral Density (PSD), from which the very low frequency (VLF), low frequency (LF), and high frequency (HF) components were extracted, along with the LF/HF ratio that reflects sympathovagal balance. Complementary measures such as Cardiac Autonomic Balance (CAB), Vagal-Sympathetic Evaluation (VSE), skewness, and kurtosis were also included to enhance signal characterization. 
Table~\ref{tab:feature_description} summarizes the complete set of extracted features and their physiological interpretation, forming the basis for subsequent machine learning and deep learning classification models.

\subsection{Classification Models}

Both machine learning and deep learning approaches were employed to evaluate stress prediction from ECG signals. For the machine learning framework, PSD-based features were used as input to a range of classifiers, spanning from simple distance-based and linear models to more complex tree-based and boosting methods. 
This diversity allowed for a comprehensive comparison across different levels of model complexity, capturing both the interpretability of traditional algorithms and the improved predictive ability of ensemble techniques.  In parallel, deep learning models were applied directly to the ECG signals to learn discriminative patterns without handcrafted feature engineering. 
Convolutional neural networks were used to extract local waveform features, while recurrent architectures such as LSTM were designed to capture temporal dependencies across heartbeat sequences. 
By combining handcrafted PSD features for machine learning with end-to-end learning in deep networks, the study provides a broad evaluation of classification strategies for stress detection in ECG signals.

\subsubsection{Support Vector Machine (SVM)}
The Support Vector Machine is capable of handling both linear and non-linear classification tasks by applying kernel functions such as polynomial, sigmoid, and Gaussian radial basis function (RBF). To ensure optimal performance, the most suitable kernel and hyperparameters were identified through a systematic parameter search using \texttt{GridSearchCV}.

\subsubsection{K-Nearest Neighbors (KNN)}
The K-Nearest Neighbors algorithm assigns a class label to a sample by considering the majority vote among its $k$ closest neighbors, with similarity often measured by Euclidean distance or similar distance metrics. The ideal value of $k$ was determined through cross-validation combined with grid search.

\subsubsection{Decision Tree (DT)}
Decision Trees operate by recursively splitting the dataset into subsets based on criteria such as entropy or Gini impurity, constructing a tree structure that makes classification decisions. Important hyperparameters, including maximum depth and split rules, were fine-tuned using \texttt{GridSearchCV} to enhance performance.

\subsubsection{Random Forest (RF)}
Random Forest is an ensemble method that constructs multiple decision trees by sampling features and instances randomly, and aggregates their predictions to improve stability and accuracy. Key parameters, such as the number of estimators and maximum tree depth, were optimized using grid search.

\subsubsection{Linear Discriminant Analysis (LDA)}
Linear Discriminant Analysis reduces dimensionality by projecting data into a space that maximizes the separation between classes relative to the variation within each class. This allows effective classification when the relationship between classes can be approximated linearly.

\subsubsection{XGBoost}
XGBoost, an implementation of gradient boosting, builds decision trees sequentially, with each successive tree correcting the residuals from the previous stage. By carefully adjusting learning rates and other hyperparameters, the algorithm achieves low bias and variance, resulting in strong predictive performance.

\subsubsection{LightGBM}
Light Gradient Boosting Machine (LightGBM) is a gradient boosting technique designed for efficiency and speed, using leaf-wise tree growth and histogram-based splitting. It naturally handles categorical variables but requires careful tuning to avoid overfitting due to its aggressive splitting strategy.

\subsubsection{CatBoost}
CatBoost is a boosting method optimized for categorical features and designed to reduce overfitting. It applies ordered boosting and constructs symmetric trees to accelerate training. CatBoost achieves high predictive accuracy with minimal feature preprocessing and hyperparameter adjustments.

\subsubsection{Convolutional Neural Network (CNN)}
CNNs are deep neural networks well-suited for spatial and visual data. They use convolutional layers to learn local features, pooling layers for dimensionality reduction, and fully connected layers for classification. CNNs are widely applied in image recognition, object detection, and video analysis.

\subsubsection{Long Short-Term Memory (LSTM)}
LSTMs are recurrent neural networks designed to capture temporal dependencies in sequential data. With memory cells and gating mechanisms, they overcome the vanishing gradient issue in standard RNNs and are widely used in time-series forecasting, speech recognition, and natural language processing.

\begin{figure}[!t]
    \centering
    \includegraphics[width=0.48\textwidth]{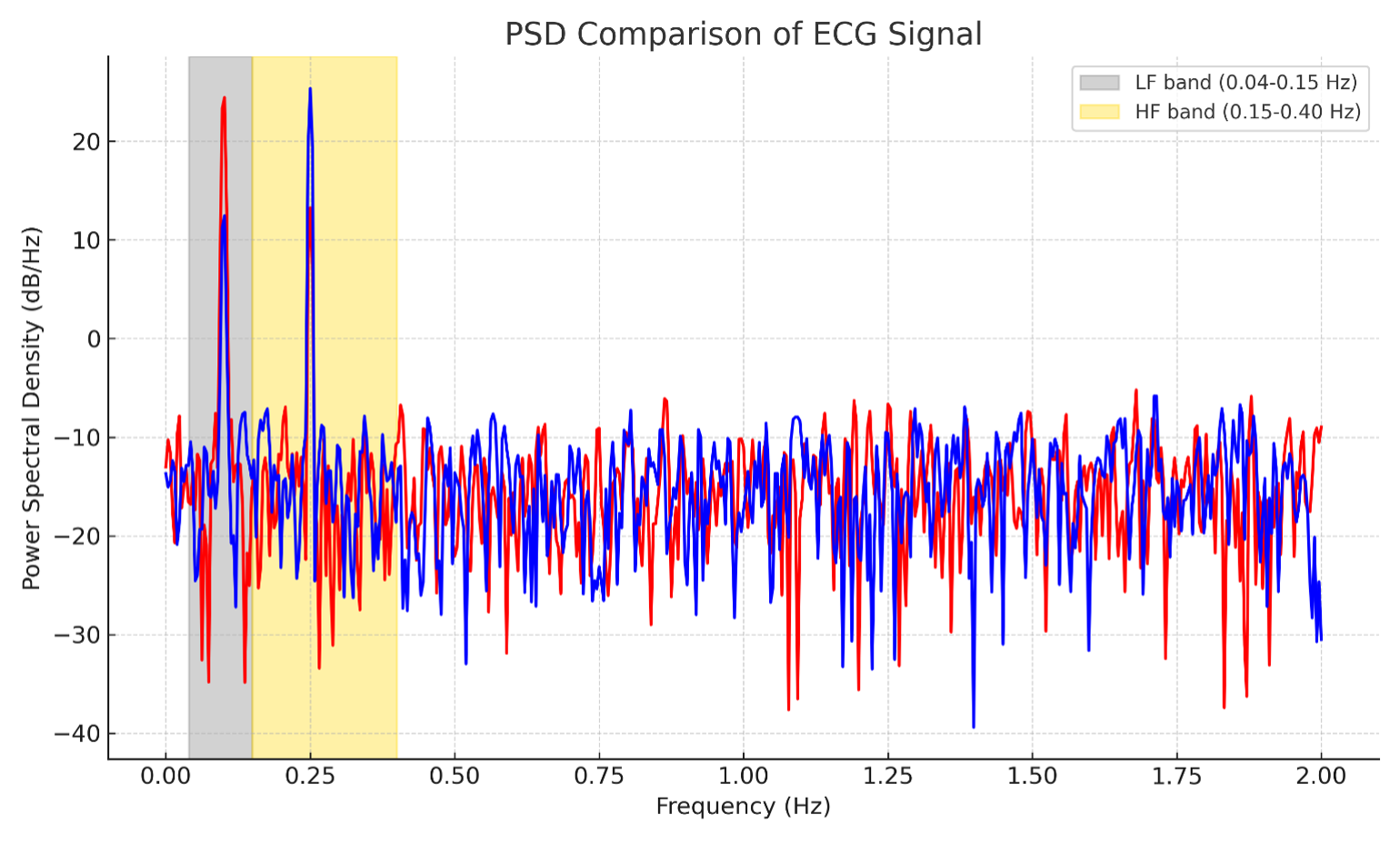}
    \caption{PSD of ECG signals with stressed (red) and non-stressed (blue) cases. LF band is marked in gray, HF band in yellow.}
    \label{fig:psd}
\end{figure}

\subsection{Evaluation Protocol}

To evaluate the classification framework, the dataset was divided into training and testing subsets with proportional representation of stressed and non-stressed samples across all activities. This stratification ensured balanced evaluation and avoided bias toward a single condition.  Performance was measured using four standard metrics: accuracy, precision, recall, and F1-score. Accuracy reflects overall correctness, precision measures the reliability of stress predictions, and recall evaluates sensitivity to true stress cases. Since precision and recall often trade off, the F1-score was included as a balanced indicator under imbalanced conditions. In addition, confusion matrices were generated to visualize prediction errors and better understand model behavior, such as whether stress was frequently confused with non-stress. The same evaluation protocol was applied consistently to all classifiers, ensuring fair comparison and enabling insights into their relative strengths for stress prediction in lifestyle contexts.

\begin{table}[htbp]
\centering
\caption{Performance comparison of classifiers for stress prediction}
\label{tab:classifier_performance}
\begin{tabular}{|l|c|c|c|c|}
\hline
\textbf{Classifier} & \textbf{Precision} & \textbf{Recall} & \textbf{F1 Score} & \textbf{Accuracy} \\
\hline
KNN       & 0.67 & 0.70 & 0.69 & 0.69 \\
SVM       & 0.75 & 0.78 & 0.76 & 0.76 \\
LDA       & 0.73 & 0.79 & 0.75 & 0.75 \\
DT        & 0.80 & 0.82 & 0.82 & 0.83 \\
RF        & 0.86 & 0.88 & 0.86 & 0.86 \\
XgBoost   & 0.85 & 0.88 & 0.85 & 0.86 \\
LightGBM  & 0.83 & 0.87 & 0.87 & 0.87 \\
CatBoost  & 0.88 & 0.91 & 0.88 & 0.90 \\
CNN       & 0.61 & 0.65 & 0.61 & 0.64 \\
\textbf{LSTM} & \textbf{0.91} & \textbf{0.93} & \textbf{0.92} & \textbf{0.94} \\
\hline
\end{tabular}
\end{table}

\section{Result and Discussion}

This section presents the experimental findings obtained from the proposed methodology. 
The results are organized into three parts. 
First, a Power Spectral Density (PSD) analysis is carried out to illustrate the frequency-domain differences between stressed and non-stressed ECG signals. 
Next, the performance of machine learning and deep learning classifiers is evaluated using accuracy, precision, recall, and F1-score, supported by visual comparisons and confusion matrices. 
Finally, a comparative study highlights the relative strengths of traditional ensemble-based classifiers and advanced deep learning approaches, demonstrating the effectiveness of the proposed framework in predicting stress under varied conditions.

\subsection{PSD Analysis}

Figure~\ref{fig:psd} presents the Power Spectral Density (PSD) analysis of ECG signals under stressed and non-stressed conditions. The red curve represents stressed signals, while the blue curve corresponds to non-stressed signals. Two frequency regions are highlighted, the low-frequency (LF) band between 0.04--0.15 Hz (shaded gray) and the high-frequency (HF) band between 0.15--0.40 Hz (shaded yellow). The PSD patterns reveal clear differences between the two conditions. Stressed signals show stronger fluctuations and higher peaks in the LF band, reflecting increased sympathetic nervous system activity, which is commonly associated with stress responses. In contrast, non-stressed signals demonstrate more stable power distribution across both LF and HF bands, indicating balanced autonomic regulation with stronger parasympathetic contributions in the HF region. These spectral differences confirm the discriminative power of PSD features for stress detection, particularly through the LF/HF ratio, which increases noticeably under stress. This observation provides strong motivation for incorporating PSD-derived features into machine learning and deep learning models for robust stress prediction.

\subsection{Performance of the Models}

\begin{figure}[!t]
    \centering
    \includegraphics[width=0.48\textwidth]{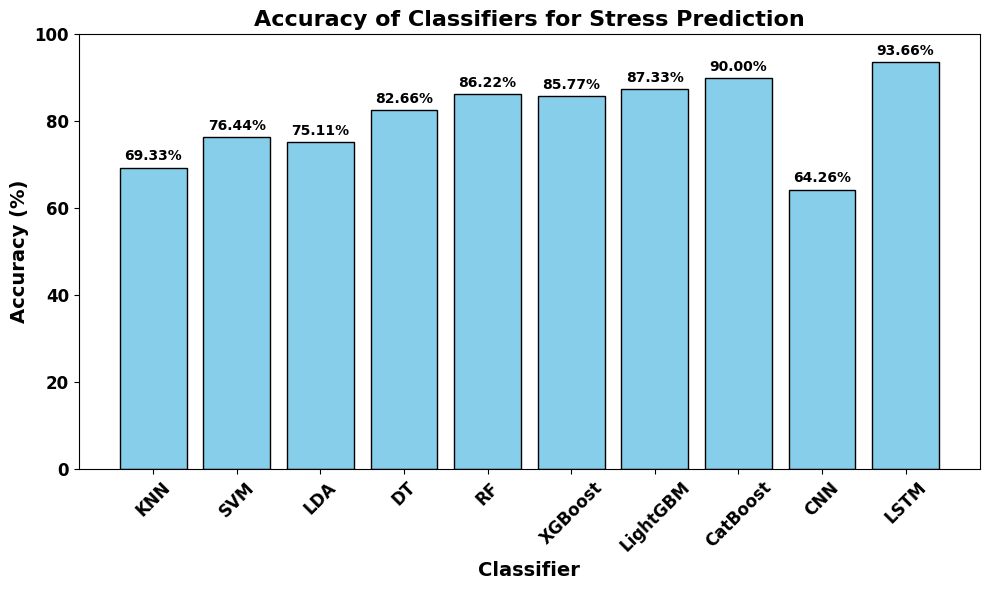}
    \caption{Accuracy of classifiers for stress prediction. LSTM achieved the highest accuracy (93.66\%), followed by CatBoost (90\%), while CNN showed the lowest performance.}
    \label{fig:accuracy_classifiers}
\end{figure}

\begin{figure}[!t]
    \centering
    \includegraphics[width=0.48\textwidth]{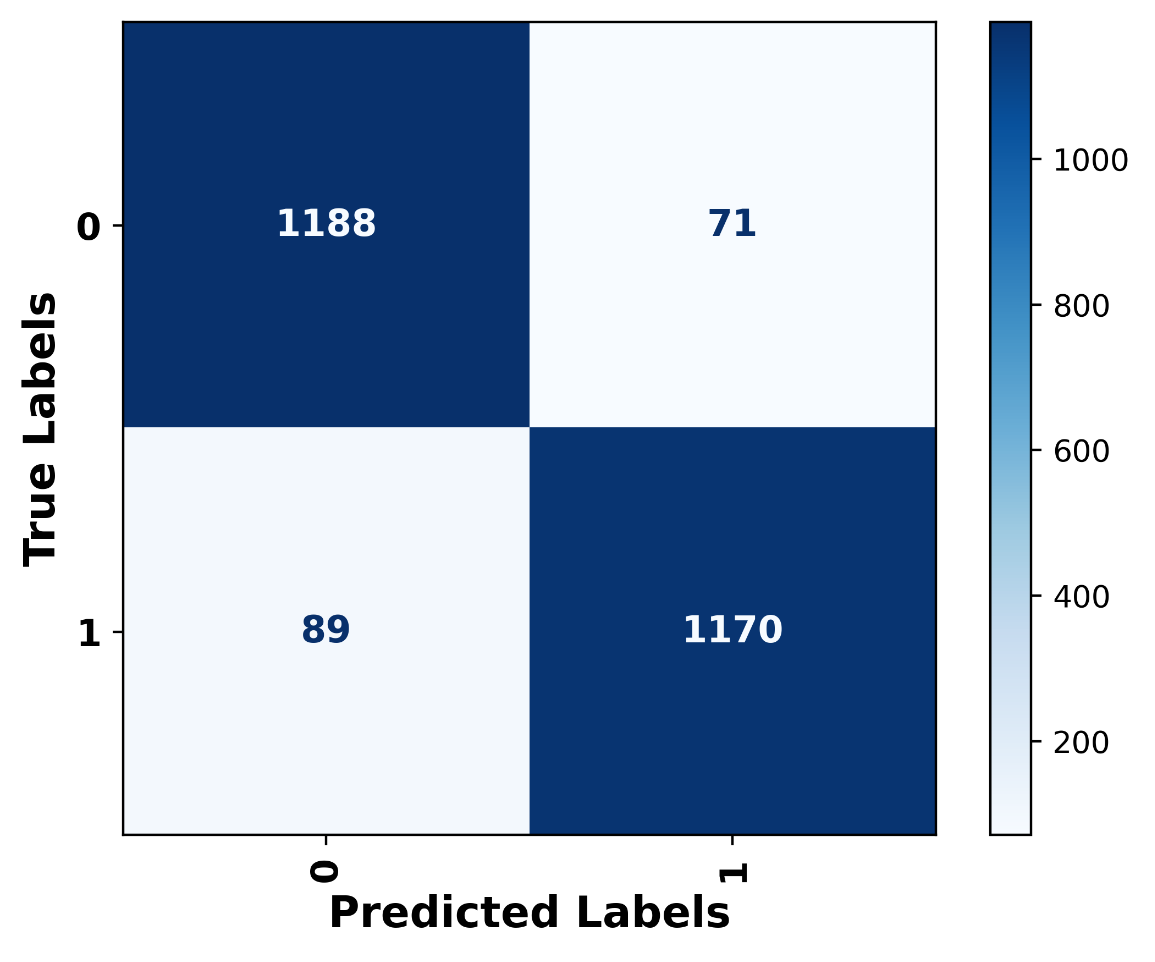}
    \caption{Confusion matrix of the LSTM classifier showing predicted vs. true labels. 
    The model demonstrates strong performance with high correct classifications in both classes.}
    \label{fig:lstm_confusion}
\end{figure}

The performance comparison of all classifiers is presented in Table~\ref{tab:classifier_performance}. 
Among the traditional machine learning models, ensemble methods such as Random Forest, XGBoost, LightGBM, and CatBoost achieved stronger results compared to simpler classifiers like KNN, SVM, and LDA. 
CatBoost obtained the highest accuracy among the machine learning models with 0.90, supported by high precision and recall values, showing its effectiveness in handling PSD-based features. 
Figure~\ref{fig:accuracy_classifiers} further illustrates the accuracy of each classifier, highlighting the superiority of deep learning methods. 
The LSTM network achieved the best overall performance with an accuracy of 0.94 and an F1 score of 0.92, outperforming CNN, which showed the lowest performance with 0.64 accuracy. 
To validate these results, Figure~\ref{fig:lstm_confusion} shows the confusion matrix of the LSTM classifier, where the model correctly classified the majority of both stressed and non-stressed samples, with only a small number of misclassifications. 
These results confirm that while boosting-based machine learning models provide competitive accuracy, deep learning models, particularly LSTM, offer the most robust solution for stress prediction from ECG signals by effectively capturing temporal dynamics.

\subsection{Discussion}

The results show that the proposed approach can effectively predict stress from ECG signals. 
The PSD analysis revealed clear differences between stressed and non-stressed conditions, with stressed signals showing stronger low-frequency power and higher LF/HF ratios that indicate greater sympathetic activity. Classifier evaluation demonstrated that ensemble-based models such as CatBoost achieved strong performance, reaching an accuracy of 0.90. Deep learning models provided even better results, with LSTM achieving the highest accuracy of 0.94 along with balanced precision, recall, and F1-score. 
The confusion matrix confirmed that LSTM correctly classified most stressed and non-stressed samples with few errors. These findings suggest that while boosting methods are effective with PSD-based features, LSTM offers a more reliable solution by capturing temporal dynamics in the ECG signals, making it the most suitable model for stress prediction in this study.

\section{Conclusion}

This paper presented a framework for stress prediction from ECG signals across different lifestyle situations. 
The analysis of Power Spectral Density (PSD) features confirmed that stressed and non-stressed states can be distinguished in the frequency domain, with stressed signals showing stronger low-frequency activity and higher LF/HF ratios. Machine learning models using PSD features achieved strong performance, particularly ensemble-based methods such as CatBoost, which reached 90\% accuracy. Deep learning models provided further improvements, with LSTM achieving the highest accuracy of 94\% along with balanced precision, recall, and F1-score. The results indicate that while boosting methods effectively capture handcrafted features, recurrent architectures such as LSTM are better suited for modeling temporal dynamics in ECG signals.  Future work will focus on expanding the dataset to include more subjects and longer monitoring periods, as well as incorporating multimodal physiological signals such as PPG and EMG. Another important direction is the development of lightweight models for wearable deployment, enabling real-time stress monitoring in everyday life. These advances can support reliable stress detection systems that contribute to better health management and early intervention in stress-related conditions.

\bibliography{references}

\end{document}